\documentclass[traditabstract]{aa} 
\usepackage{natbib}
\usepackage{longtable,lscape}
\usepackage{graphicx}
\usepackage{txfonts}
\begin{document}
   \title{Optically faint X-ray sources in the CDFN: Spitzer constraints}


   \author{E. Rovilos\inst{1}
          \and
          I. Georgantopoulos\inst{2,3}
          \and
          A. Akylas\inst{3}
          \and
          S. Fotopoulou\inst{4,5}
          }
   \institute{Max Planck Institut f\"{u}r Extraterrestrische Physik,
              Giessenbachstra\ss e, 85748 Garching, Germany
          \and
              Osservatorio Astronomico di Bologna/INAF, Via Ranzani 1, 40127
              Bologna, Italy
          \and 
              Institute of Astronomy and Astrophysics, National Observatory of
              Athens, I. Metaxa \& V. Pavlou str., Palaia Penteli, 15236 Athens,
              Greece
          \and
              Max Planck Institut f\"{u}r Plasma Physik, Boltzmannstra\ss e 2,
              85748 Garching, Germany
          \and
              Technische Universit\"{a}t M\"{u}nchen, Physik-Department, James
              Frank Stra\ss e, 85748 Garching, Germany 
             }

   \date{}

 \abstract{We investigate the properties of the most optically faint sources in the GOODS-N area
  ($R_{AB} > 26.5$). Such extremely optically faint 
 populations present  an uncharted territory despite the fact that they 
 represent an appreciable fraction  of the X-ray sources in the GOODS-N field.  
  The optically faint sources are believed to contain either red AGN at 
  moderate redshifts or possibly QSO at very high redshift. 
  We compile our sample by first finding   
  the $3.6\mu m$ IRAC counterparts of the X-ray sources 
  and in turn by searching for the optical counterparts of the IRAC sources.
 35 sources do not have counterparts in the $R$-band Subaru optical images. 
   Of these, 18 have {\it HST}
  ACS counterparts while the remaining have no optical counterparts.   
  The vast majority of our 35 sources are classified as Extremely Red Objects
 (EROs) on the basis of their $V_{606}-K_{\rm S}$ lower limits.   
 Their derived photometric redshifts  show that these       
  populate moderate redshifts (median $z\sim 2.8$), 
 being at markedly different redshifts from the already 
 spectroscopically identified population which peaks at $z \sim 0.7$.
      The {\it Spitzer} IRAC mid-IR colours of the
 sources which have no {\it HST} counterparts 
    tend to lie within the mid-IR colour diagram AGN ``wedge'', 
 suggesting either QSO, ULIRG (Mrk231) templates
  or early-type galaxy templates at $z>3$.
 A large fraction of our sources (17/35), regardless of whether they 
 have {\it HST} counterparts, can be classified as mid-IR bright/optically faint 
 sources (Dust Obscured Galaxies) a class of sources which is believed to include many  
 heavily absorbed AGN.  The co-added X-ray spectrum of the optically faint sources is very flat
  having a spectral index of $\Gamma \approx0.87$, significantly flatter than the spectrum of 
   the X-ray background.      
  The optically faint ($R>26.5$) X-ray sources constitute more than 50 per cent of the 
   total X-ray population at redshifts $z>2$ bearing important implications 
 for the luminosity function and its evolution; considering X-ray sources with
$2<z<4$ we find good agreement with a modified Pure Luminosity Evolution (PLE)
model.
    \keywords{Galaxies: active -- Galaxies: high-redshift -- X-rays: galaxies -- Infrared: galaxies}
               }

   \maketitle
%

\section{Introduction}

X-ray surveys provide the most efficient method for detecting AGN (Brandt \&
 Hasinger 2005). 
 This is because X-ray wavelengths can penetrate large amount of 
 interstellar gas and reveal
the AGN even in very obscured systems.
The deepest X-ray surveys to date detect a large number of sources down to a
flux of $\sim2\times10^{-17}$\,erg\,cm$^{-2}$\,s$^{-1}$ in the (0.5-2.0)\,keV
band \citep{Alexander2003,Luo2008}. The vast majority of them are
AGN \citep{Bauer2004} with a surface density of about 5000 sources per square
degree. Optical follow-up observations have identified a large fraction of them
\citep{Barger2003,Capak2004,Trouille2008}, revealing that the peak of the
redshift distribution based on spectroscopic identifications is at $z=0.7$. 
However, a large number of the X-ray sources remain optically 
 unidentified hampering our understanding of their nature. 
 In particular, a large fraction of faint X-ray
sources ($\sim50\%$) lacks a spectroscopic identification
\citep[e.g.][]{Luo2010} and the redshift
estimate is made with photometric techniques. Still,  
 \citet{Aird2010} estimate that about one third of the X-ray sources 
 in the CDFs do not have optical counterparts down to $R_{AB}\approx 26.5$. 

The nature of these optically faint sources remains puzzling. 
Two interesting scenaria have been proposed to explain their nature. 
First, the very
faint optical emission could be the result of copious dust absorption; in the
{\it Chandra} deep fields the majority of high $f_{\rm x}/f_{\rm o}$ sources shows
clear evidence of obscuration in their individual and stacked X-ray spectra
\citep*{Civano2005}, a result which confirms previous findings
\citep{Alexander2001}. 
On the other hand, a very faint, or the lack of an optical counterpart could
indicate a very high redshift source \citep{Koekemoer2004}. In this case, the
reason of a high X-ray to optical ratio is that the optical bands are probing
bluer rest-frame wavelengths, which are more obscured or intrinsically fainter
if they fall blueward from the Lyman break. At the same time  the observed X-ray
wavelengths correspond to high energy rest-frame wavelengths which are less
prone to absorption.
\citet{Lehmer2005} used a Lyman break technique to select high redshift
galaxies in the {\it Chandra} deep fields and found 11 $B_{435}$, $V_{606}$, and
$i_{775}$ dropouts among the X-ray sources, with possible redshifts $z\gtrsim4$.

A crucial diagnostic for the nature of optically faint galaxies is their infrared
emission. The optical and ultra-violet light which is absorbed is re-emitted
in infrared wavelengths, therefore a high infrared to optical ratio can be
used as a criterion of high obscuration. For example, \citet{Houck2005} discovered a number of sources
with extremely high 24$\mu{\rm m}$ to optical luminosities
\citep[see also][]{Daddi2007}. These sources,
nicknamed DOGs (Dust Obscured Galaxies) are located at $1.5\lesssim z\lesssim 2$
\citep{Pope2008}.   A large fraction of them
is probably associated with Compton-thick QSOs
\citep{Fiore2008,Fiore2009,Treister2009,Georgantopoulos2008}.

In this paper we explore the properties of optically faint
($R_{\rm AB}\gtrsim26.5$) X-ray sources in the GOODS-N area.
  Previous studies of some of the optically faint sources 
 in this field have been 
 performed in the past; \citet{Alexander2001} studied the ``brightest'' optically
faint sources (with $I>24$) and \citet{Civano2005} investigated the properties
of Extreme X-ray Objects (EXOs), having an X-ray to optical flux ratio $>10$.
Instead, our sample focuses not only on EXOs but on all optically faint sources.
Additionally, we take advantage of the full 
 2\,Ms X-ray exposures, but more importantly we make use of the
 mid-IR {\it Spitzer} observations in order to find the true counterparts 
  of the X-ray sources.   
 We perform anew the search for optical identifications 
 of the X-ray sources in two steps. First, we   
  cross-correlate the X-ray positions with the {\it Spitzer} IRAC
3.6\,$\mu{\rm m}$ catalogue 
 using a likelihood ratio technique. Then we cross-correlate the IRAC positions 
 with the \citet{Capak2004} $R$-band optical observations. We define the optically 
 faint sources as those with $R_{\rm AB}\gtrsim26.5$, i.e. beyond the flux limit 
 of this catalogue. Then we look for optical counterparts in the fainter {\it HST} 
 observations. After deriving photometric redshifts 
 (section 4) we study their  X-ray  (section 5)
 as well as their mid-IR properties (section 6).  


\section{X-ray, optical and IR data and counterpart association}

We use the 2\,Ms CDFN catalogue of \citet{Alexander2003}, with a sensitivity
of $2.5\times 10^{-17}\,{\rm erg}\,{\rm cm}^{-2}{\rm s}^{-1}$ in the
0.5-2.0\,keV band and $1.4\times 10^{-16}\,{\rm erg}\,{\rm cm}^{-2}{\rm s}^{-1}$
in the 2.0-8.0\,keV band. The infrared ({\it Spitzer}) data come from the
{\it Spitzer}-GOODS legacy programme \citep{Dickinson2004}. This includes observations
in the mid-infrared with IRAC and MIPS which cover most of the
CDFN area. Typical sensitivities of these observations are $0.3\,\mu{\rm Jy}$
and $80\,\mu{\rm Jy}$ for the IRAC-$3.6\,\mu{\rm m}$ and MIPS-$24\,\mu{\rm m}$
bands respectively.
Optical coverage of the GOODS-North field is available both with ground-based
(Subaru-SuprimeCam) and space observations ({\it HST}-ACS). The Subaru sensitivity is
26.5\,mag(AB)
in the $R$ band \citep{Capak2004} and the {\it HST} is 27.8\,mag(AB) in the
$z_{850}$ band \citep{Giavalisco2004}. We also make use of near-infrared
($K_S$-band) observations with Subaru-MOIRCS, reaching a limit of 23.8\,mag(AB)
\citep{Bundy2009}.

\begin{figure}
  \resizebox{\hsize}{!}{\includegraphics{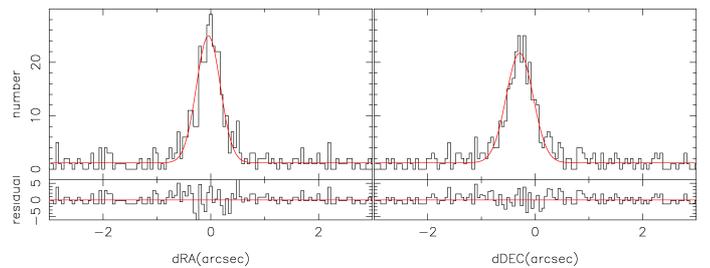}}
  \caption{Distance in RA and DEC between X-ray and infrared sources. There
           is a significant shift in declination between the two catalogues
           which we correct for before making the correlation.}
  \label{rel_astro}
\end{figure}

The X-ray catalogue \citep{Alexander2003} has 503 sources detected in one of
the hard (2.0-10.0\,keV) soft (0.5-2.0\,keV), or full (0.5-10.0\,keV) bands.
Of these sources, 348 fall into the region covered by the IRAC observations.
We make an initial simple cross-correlation of the positions of the sources
of the two catalogues to check their relative astrometry. We find good
agreement in the RA axis ($\rm{<RA(Xray)-RA(IR)>=-0.05}$, $\sigma=0.31$),
but there is a significant difference in the astrometry in the DEC axis
($\rm{<DEC(Xray)-DEC(IR)>=-0.29}$, $\sigma=0.35$, see Fig.\,\ref{rel_astro});
we update the positions of the IR sources accordingly before proceeding to the
search of IR counterparts to the X-ray sources.

\begin{figure}
  \resizebox{\hsize}{!}{\includegraphics{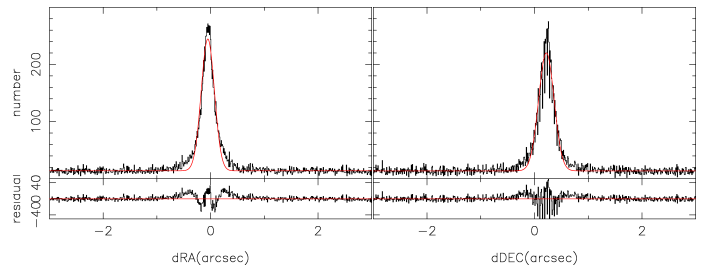}}
  \caption{Same as Fig.\,\ref{rel_astro} but for the {\it Spitzer} (without including
           the correction introduced to match the X-ray astrometry) and optical
           catalogues. The shift in declination is again corrected for before
           correlating.}
  \label{rel_astro_2}
\end{figure}

We use ``likelihood ratio'' method \citep{Sutherland1992} to find
the counterparts: we calculate the likelihood ratio ($LR$) of an
infrared source being a real counterpart of an X-ray source as:
\[LR=\frac{q(m)f(x,y)}{n(m)}\]
where $q(m)$ is the expected magnitude distribution of real counterparts,
$f(x,y)$ is the probability distribution function of positional errors, a
gaussian in this case, and $n(m)$ is the magnitude distribution of background
objects. We calculate the $q(m)$ function by subtracting the magnitude
distribution of the background (IR) sources we expect to find within the search
radius near the target (X-ray) sources from the magnitude distribution of all
the initial counterparts. The normalization of $q(m)$ is done using
\[\int_{-\infty}^{M_{\rm lim}}q(m){\rm d}m=Q(M_{\rm lim})\]
where $Q(M_{\rm lim})$ is the probability that the IR counterpart is brighter
than the magnitude limit $M_{\rm lim})$; in practice it is the final fraction of
X-ray sources with an IR counterpart. Given that the positional offsets of
the X-ray sources in the CDFN at large off-axis angles can be as large
as $1\arcsec-2\arcsec$ \citep{Alexander2003}, we use a large initial search
radius for the X-ray - IR counterparts (4\arcsec).

The choice of an optimum likelihood ratio cutoff is a compromise between
the reliability of the final sample and its completeness. The reliability
of a possible counterpart is defined as:
\[R_i=\frac{LR_i}{\sum LR_j+(1-Q(M_{\rm lim}))}\]
where $j$ refers to the different IR counterparts to a specific X-ray
source. We chose the likelihood ratio cutoff which maximizes the sum of
completeness and reliability of the matched catalogue in the same manner as
\citet{Luo2010}. The completeness is defined as the ratio of the sum of the
reliabilities of the counterparts with $LR>LR_{\rm lim}$ with the number of the
X-ray sources in the area mapped by IRAC, and the reliability is defined as the
mean reliability of counterparts with $LR>LR_{\rm lim}$. In the X-ray - IR case
$LR_{\rm lim}=0.15$, which gives a reliability of 99.2\%.

We detect 330 IR counterparts with $LR>0.15$, we then optically inspect the
positions on the IRAC images of the X-ray sources lacking a counterpart and find
12 cases where the IR source is clearly visible but blended with a nearby source
in a way that the source extraction algorithm could not distinguish them. Adding
these cases, the final number of counterparts is 342 out of the 348 X-ray sources
in the IRAC area. Assuming that 0.8\% of the counterparts are spurious, the
final efficiency is 97.5\%.

In order to look for optical counterparts to the X-ray sources we use the
positions of their IR counterparts. The positional accuracy of IRAC is better
than that of {\it Chandra}, moreover the NIR is more efficient in detecting AGN than
the optical emission, due to the reprocessing of the ionizing radiation of the
AGN through circumnuclear and interstellar dust to IR wavelengths. As a result,
the efficiency of IRAC in finding counterparts for the X-ray sources ( 97.5\%)
is higher than this of the optical survey (78.1\% if we follow the procedure
described above for the Subaru-$R$ catalogue), in spite of the optical being
deeper than IRAC \citep[see also][for the CDFS case]{Luo2010}.

The optical catalogue of \citet{Capak2004} has 47450 sources
detected in the $R$ band, and of those 14763 fall into the area sampled by IRAC.
Before searching for optical counterparts, we check again the relative
astrometry of the optical and (uncorrected) IR catalogues. We find again a similar astrometry
difference in the DEC axis (${\rm <DEC(opt)-DEC(IR)>=-0.22}$, $\sigma=0.20$,
Fig\,\ref{rel_astro_2}), while the RA positions are well within the error
(${\rm <RA(opt)-RA(IR)>=0.05}$, $\sigma=0.17$). We again correct the IR
catalogue to match the optical positions.

We use the likelihood ratio method to select the optical counterparts
to the IR sources. Given the smaller PSF of the optical images we use a smaller
initial search radius (3\arcsec) and follow the same procedure to select the
optimum likelihood ratio limit. With $LR_{\rm lim}=0.25$ which this time gives a
mean reliability of 98.4\% we find optical counterparts for 8437 of the 10595
IRAC sources. The recovery rate is 78.0\%. We perform also an independent
search for optical counterparts of X-ray sources which do not have an infrared
counterpart, or the latter is a merged source as described above, and find
a secure optical counterpart for 12 such cases; 11/12 of the ``confused'' IR
sources have an optical source related to the X-ray position, while 1/6 of the
IRAC non-detections is detected in optical. 
Finally, we optically inspect  the Subaru R-band images of X-ray source lacking
an optical counterpart and find that in 17 the optical counterpart is clearly
visible but with no identification in the catalogue of \citet{Capak2004}. Two
common reasons for that are that the source is saturated, and therefore it does
not have reliable photometry, and that it lies close to a very bright source
and therefore it is not detected by the source extracting algorithm.

The X-ray, optical, and infrared properties of all the sources of the common
CDFN-IRAC area can be found in
{\tt ftp.mpe.mpg.de/people/erovilos/CDFN\_IR\_OPT/}


\section{Sample selection}
\label{sample_selection}

\addtocounter{table}{1} 

In this study we are interested in X-ray sources which are too faint to
be detected by typical ground-based optical surveys, like that of
\citet{Capak2004}. Such sources represent a sizable fraction of the overall
X-ray population ($\sim 15\%$). They are generally not covered with
spectroscopic surveys, which are typically magnitude limited with
$R_{\rm lim}<26.5$ and often do not even have photometric redshifts. Such cases
have been studied before in detail \citep[e.g.][]{Alexander2001,Mainieri2005};
here we approach them using their infrared properties and X-ray spectra.

There are 310 X-ray sources within the common area of all the
surveys used in this study (2\,Ms CDFN - Subaru-Suprime(optical) -
Subaru-MOIRCS($HK'$) - GOODS-ACS - GOODS-IRAC - GOODS-MIPS). Of these 310
sources, 42 are too faint to be detected in the $R$-band by Subaru-Suprime and
are assumed to have $R_{\rm AB}>26.5$. No $R$-detected CDFN source has
$R_{\rm AB}>26.5$. Because in this study we are based on the infrared properties
to examine the nature of optically faint sources, we exclude 7 of the 42 sources
from our studied sample for the following reasons: 5 are not detected with IRAC,
and 2 are blended with nearby sources, so their infrared photometry is not
reliable. Our final sample consists of 35 sources, whose X-ray, optical, and
infrared properties are listed in Tab.\,\ref{sample}. Among these sources we
expect a mean number of 0.35 spurious encounters (practically none), as the mean
reliability of the IRAC counterparts for these 35 cases is 99.0\%.

For these 35 sources we search the $K_S$ catalogue of \citet{Bundy2009} for
counterparts. These Subaru-MOIRCS images cover an area slightly smaller than
IRAC and we limit our field to the common area. As the number of sources is
small, we look for $K_S$-band counterparts by eye; we can this way easily
distinguish any $K_S$-band sources being physically related to a nearby IRAC
source. We find a $K_S$ counterpart for 29/35 sources; their distances from the
IRAC sources are all $<1.4\arcsec$, while for the 6 non-detections the distance
of the nearest $K_S$ source is always $>3.9\arcsec$. We then search the
$HK'$-band images from the UH-2.2\,m telescope \citep{Capak2004} to detect any
sources bright enough in the $HK'$ band but not detected in $R$. Using
sextractor \citep{Bertin1996}, we identify as a source 4 adjacent pixels with
fluxes above 1.2 times the local background rms. We find an $HK'$ detection for
19 of the sources in our sample. The Subaru-MOIRCS area has very deep optical
imaging with the {\it HST}-ACS as part of the GOODS survey. The catalogues
\citep{Giavalisco2004} are based on detections in the $z_{850}$ band and are
publicly available. We search the sources of our sample for {\it HST}
counterparts using the likelihood ratio method, as described in the previous
section with an initial search radius of 1.5\,arcsec. We use this method
because the PSFs of {\it Spitzer}-IRAC and {\it HST}-ACS are different within a
large factor and there could be multiple ACS sources within one IRAC beam.
18/35 sources of our sample have an {\it HST}-ACS counterpart.

\begin{figure*}
  \resizebox{\hsize}{!}{\includegraphics{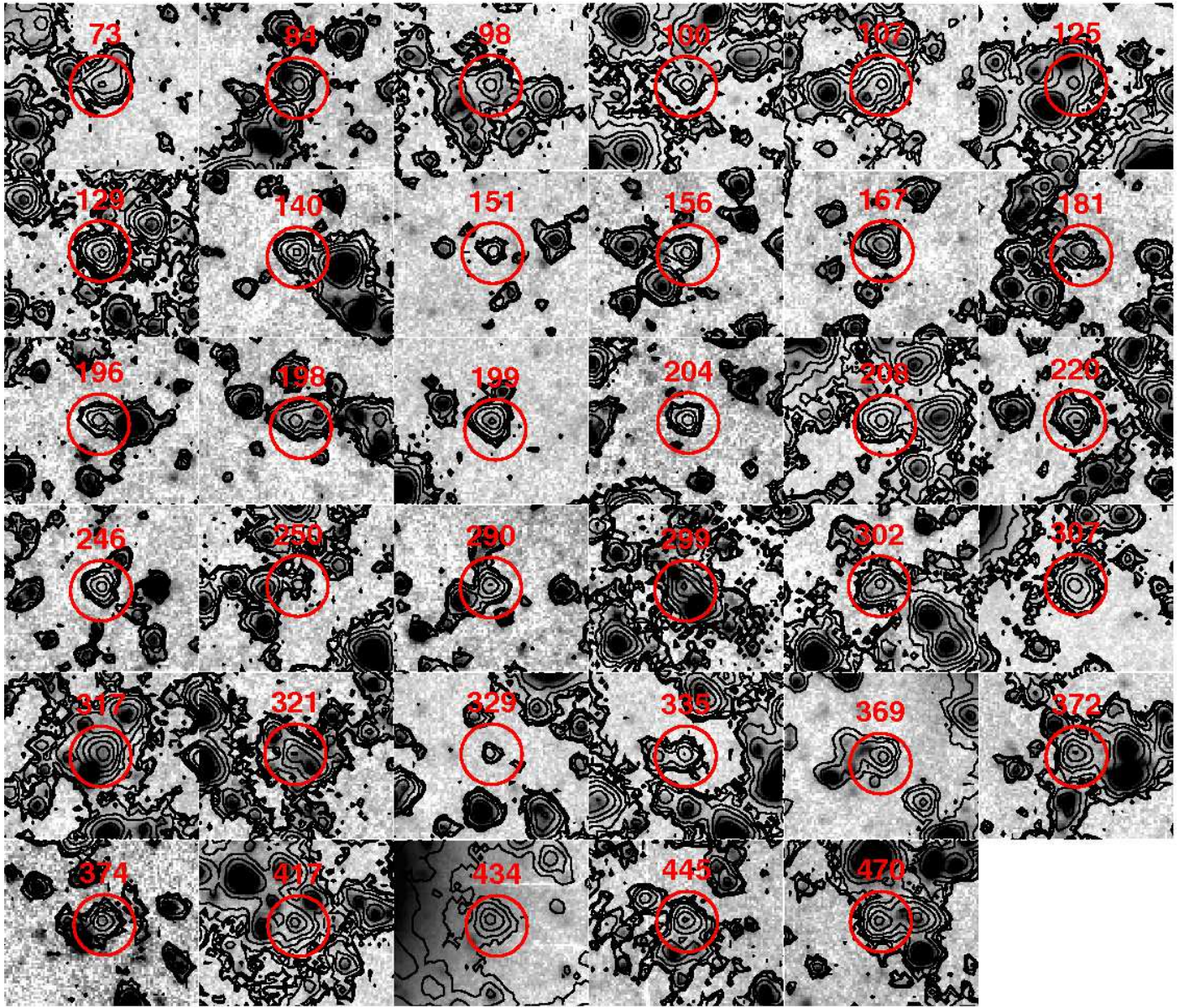}}
  \caption{Cutout images of the sources listed in Tab.\,\ref{sample}. In
           greyscale is the Subaru $R$-band from \citet{Capak2004}. Large
           circles mark the position of the X-ray sources with a 4\arcsec
           radii and contours represent the IRAC $3.6\,\mu{\rm m}$ flux.}
  \label{opt_positions}
\end{figure*}

We note here that some of the sources in Tab.\,\ref{sample} appear with full
optical
photometry in \citet{Barger2003}, some of them even with a photometric redshift.
In cases of X-ray sources lacking an optical counterpart, like the ones in
Tab.\,\ref{sample}, \citet{Barger2003} measured the optical fluxes directly
from the Subaru images using a 3\arcsec diameter aperture centered on the
position of the X-ray source. In doing so there is a high probability that
light from a neighboring source enters the aperture, moreover centering on the
X-ray positions causes a loss in positional accuracy, which is essential when
measuring the flux of a ``non-visible'' source. Fig.\,\ref{opt_positions} shows
the $R$-band images of the 35 sources of Tab.\,\ref{sample}, with 4\arcsec-radii
circles on the X-ray positions (as large as the initial search radius) as well
as contours representing the IRAC flux. We can see that in some cases the X-ray
and IRAC positions differ significantly and that the 1.5\arcsec aperture centered
on the X-ray position would often be contaminated by nearby optical sources.

\begin{figure}
  \resizebox{\hsize}{!}{\includegraphics{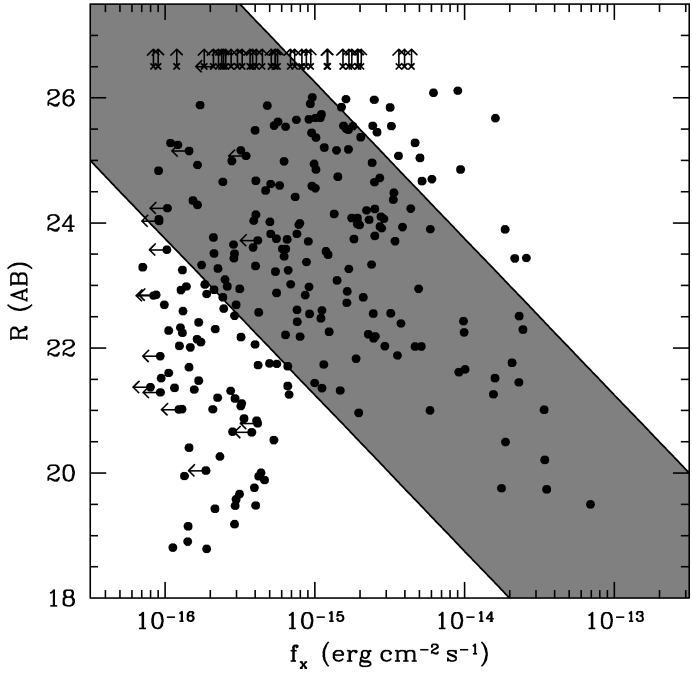}}
  \caption{Optical versus X-ray flux for all the X-ray sources in the common
           area (see text). Crosses mark the sources of Tab.\,\ref{sample} and
           the shaded area marks $\log(f_{\rm x}/f_{\rm opt})=0\pm 1$ where the
           bulk of AGN are expected.}
  \label{fxfR}
\end{figure}

\begin{figure}
  \resizebox{\hsize}{!}{\includegraphics{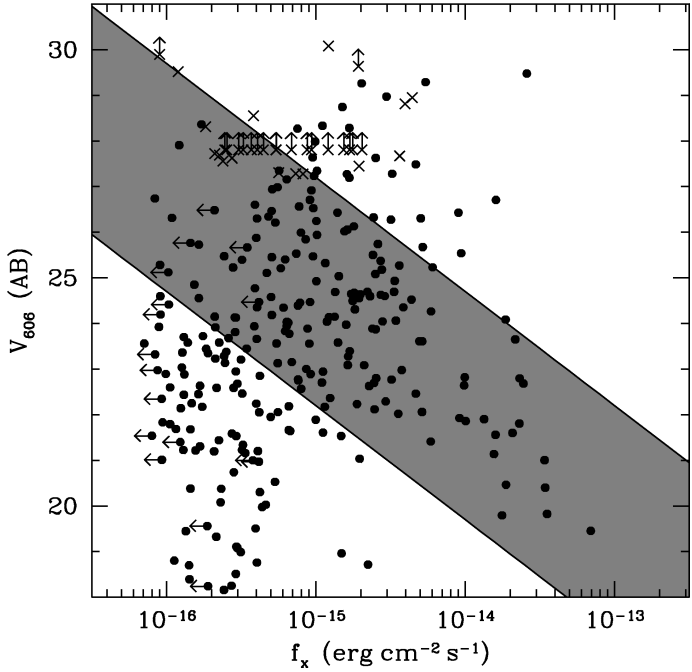}}
  \caption{Same as Fig.\,\ref{fxfR} but for the {\it HST}-ACS $V_{606}$ optical flux.
           The $V_{606}>27.8$ upper limit for non detection is likely to be an
           under-estimation (see text).}
  \label{fxfV}
\end{figure}

The sample presented in Tab.\,\ref{sample} has both similarities and differences
with the high X-ray to optical ratio sample \citep{Civano2005}, and
the optically faint sample presented by \citet{Alexander2001}. In
Fig.\,\ref{fxfR} we plot the $R$ magnitudes of the optical counterparts against
the (0.5-8.0)\,keV flux as filled circles and the sources in our sample
(all lower limits) as crosses. The shaded area is the
$\log(f_{\rm x}/f_{\rm opt})=0\pm 1$ area where the bulk of AGN are expected
\citep[e.g.][]{Elvis1994}. Our sources have, on average, a higher
$f_{\rm x}/f_{\rm opt}$ value than the bulk of the AGN population, but could not
be securely considered as high $f_{\rm x}/f_{\rm opt}$
\citep[with a value $>10$,][]{Koekemoer2004}. The {\it HST} $V_{606}$ magnitudes
provide a clearer picture. In Fig.\,\ref{fxfV} we plot the $V_{606}$ magnitudes
against the X-ray flux keeping the same symbols as in Fig.\,\ref{fxfR}. The
$V_{606}$ magnitudes for non-detected sources are assumed to be $V_{606}>27.8$,
the same as the detection threshold in the $z_{580}$ band. The true limits
however are likely to be higher, as the typical $V_{606}-z_{850}$ colour of AGN 
hosts is $>0.5$ and increasing with redshift \citep{Sanchez2004}; the $2\sigma$
limits of sources 246 and 317 which are detected in the $z_{580}$ band and not
in the $V_{606}$ band are 29.6 and 29.9 respectively.
In Fig.\,\ref{fxfV} we see that the sources of
Tab.\,\ref{sample} are clearly relatively high $f_{\rm x}/f_{\rm opt}$, all having
$\log(f_{\rm x}/f_{\rm opt})>0.52$. However, compared to \citet{Civano2005}, this
is neither fully a high $f_{\rm x}/f_{\rm opt}$ sample, nor a complete one, as
many sources with $\log(f_{\rm x}/f_{\rm opt})>1$ are not included.
Moreover, the high $f_{\rm x}/f_{\rm opt}$ sample of \citet{Civano2005} includes
much brighter sources, as bright as $R\sim 23$.

Compared to \citet{Alexander2001}, our sample probes much fainter sources;
the \citet{Alexander2001} sample has a cut at $I=24$. As a result, their sources
have relatively low X-ray to optical flux ratios, in many cases
$\log(f_{\rm x}/f_{\rm opt})<0$.

\section{Photometric redshifts}
\label{photozs}

 We use the EAZY code \citep*{Brammer2008} to
calculate photometric redshifts. We used 4 ACS ($B_{435}$, $V_{606}$,
$i_{775}$, and $z_{850}$) optical, $HK'$ (when available), $K_S$ and 2 IRAC
bands ($3.6\,\mu m$, $4.5\,\mu m$) to constrain the photometric redshifts.
We do not use the $5.8\,\mu m$ and $8.0\,\mu m$ bands because they are more
sensitive to the properties of (interstellar and circumnuclear in AGN cases
like here) dust, something which would add extra parameters which would have
to be considered in the template fitting \citep[see][]{RowanRobinson2008}.

The results are also shown in Tab.\,\ref{sample}; the {\it HST} magnitudes
shown as lower limit are detections with a $<2\sigma$ confidence in the
respective bands, and the $2\sigma$ flux is used to calculate the magnitude.
The photometric redshifts based on these lower limits
are considered less reliable. Note that source 369, despite being 
very faint in the optical, has a reported spectroscopic
redshift of $z=2.914$ in \citet{Chapman2005}. The position of the
spectroscopic source comes from the radio (VLA - 1.4\,GHz) catalogue of
\citet{Richards2000} and agrees within 0.15\arcsec with the IRAC position.
The spectrum is typical of an AGN (C{\small IV} line) and the X-ray position is
1.1\arcsec away, so we are confident that this is the correct counterpart.
The photometric redshift derived by EAZY (2.80) is close to the spectroscopic
value.

For sources with no ACS detection, it is very challenging to derive a
photometric redshift based only on the IRAC and $K_S$ bands. In these cases we
are forced to use all IRAC bands, which adds the extra variable of the dust
properties. A good example of poor photo-z accuracy in cases where only
mid-infrared fluxes are fitted can be found in \citet{Salvato2009}. The
redshift derived is typically $2<z<3$, but the constrain is weak and the
broad-band spectrum in some cases can be equally well fit with SEDs shifted to
$z>3.2$. The reason for that is that in cases where the IRAC SEDs are monotonic
in $f_{\nu}$ the fit cannot be easily constrained. A good photo-z estimate with
a value of $z\sim2.5$ is derived in cases of a blue [5.8]-[8.0]
colour\footnote{We define here blue as
$f_{\nu\,5.8\,\mu{\rm m}}>f_{\nu\,8.0\,\mu{\rm m}}$, corresponding to
$[5.8]-[8.0]<0.64$}, which can be fitted with the red-NIR bump of a galactic
(no AGN) SED due to moderate temperature interstellar gas
\citep[see][Figure 8]{Salvato2009}. From Tab.\,\ref{sample}, we can see that of
the 14 sources with full IRAC photometry and no ACS detection, 5
have blue [5.8]-[8.0] colour (98, 140, 156, 198, and 372). The
median photometric redshift for these five sources as derived by EAZY is
$z=2.39$, whereas the median photo-z for the red [5.8]-[8.0] colours is
$z=4.40$. The two samples define different populations in terms of photometric
redshifts with a confidence level of 99.5\,\%, according to a K-S test.
Moreover, the ``{\it HST}-detections'' and ``{\it HST}-non-detections'' populations are
also different within 98.9\,\%, having median redshifts 2.61 and 3.48
respectively.

\begin{figure}
  \resizebox{\hsize}{!}{\includegraphics{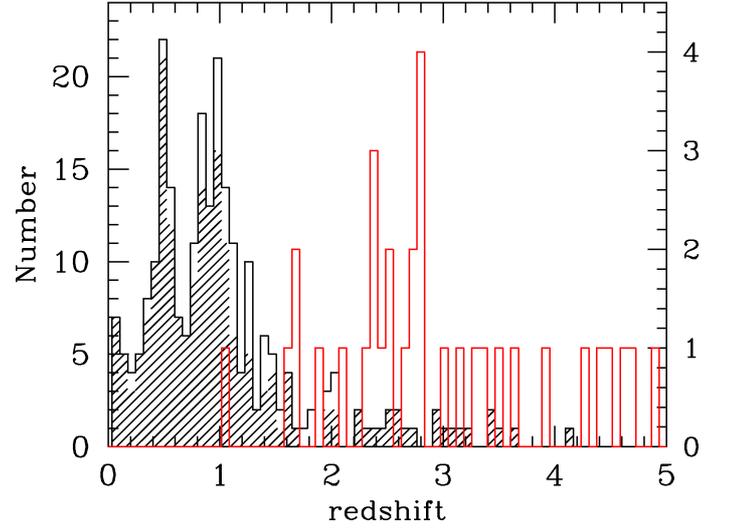}}
  \caption{Redshift distributions of CDFN sources in the common
           CDFN-GOODS-MOIRCS area. The black histogram represents all optically
           detected sources with the shaded histogram representing
           spectroscopic redshifts. The red histogram (refering to the right
           axis scale) represents the optically faint sources.}
  \label{z_dist}
\end{figure}

The redshift distribution of the sources of Tab.\,\ref{sample} is compared with
the distribution of all X-ray sources in the common CDFN-GOODS-MOIRCS in
Fig.\,\ref{z_dist}. The black histogram represents all optically detected
sources and the shaded histogram represents sources with spectroscopic
redshifts, taken from the catalogues of \citet{Barger2003} and
\citet{Trouille2008}. The red histogram (refering to the right axis scale)
represents the sources of Tab.\,\ref{sample}. The two distributions are
remarkedly different, with the sources of Tab.\,\ref{sample} being at higher
redshifts.


\section{X-ray spectral analysis}

The data have been analyzed using the CIAO v.4.1  
analysis software. The source spectra are extracted from 
circular regions with variable radii so as to include at 
least 90 per cent of the source photons in all off-axis angles. 
There are 20 X-ray observations comprising the total $\sim$2\,Ms exposure. 
We extract the source  spectrum and auxiliary files for each 
observation separately using the CIAO {\sl SPECEXTRACT} script. Then we use 
{\sl MATHPHA} {\sl ADDARF} and {\sl ADDRMF} tasks of FTOOLS to merge the 
spectral products for each source. The background files
are calculated  from source free regions in each observation 
and are again merged using the {\sl MATHPHA} tool.

We explore the X-ray properties of the 35 sources in our sample 
using the {\sl XSPEC} v12.5 package to perform X-ray spectral 
fittings. For the sources with adequate count statistics 
(net source counts $\geq 200$, 5 sources), we use the $\chi^2$ statistic
technique. 
The data are grouped to give a minimum of 15 counts per bin to ensure 
that Gaussian statistics apply. We adopt an absorbed power-law model 
and attempt to constrain the intrinsic absorption column density 
$N_{\rm H}$ (i.e., having subtracted the Galactic absorption) and the 
power-law photon index $\Gamma$. 

For the sources with limited photon statistics (net counts $<$200), 
we use the C-statistic technique \citep{Cash1979} specifically developed 
to extract spectral information from data with a low signal-to-noise 
ratio. In this case, the data are grouped to give a minimum of 1 
count per bin to avoid zero count bins. We try to constrain the 
intrinsic column densities using an absorbed power-law model 
with $\Gamma$ fixed to 1.8. 
In both cases, the spectral fittings are performed in the 0.3-8 keV 
energy band where the sensitivity of the {\it Chandra} detector is the 
highest. The estimated errors correspond to the 90 per cent 
confidence level. In Tab. \ref{xray_prop} we present the spectral 
fitting results for the 35 sources comprising our final dataset. 
 Source 107 has a very flat photon index, which is likely to be the result of a reflection-dominated
spectrum \citep[see][for a detailed analysis of the candidate Compton-thick
sources in the CDFN]{Georgantopoulos2009}. In this case the hydrogen
column density in Tab.\,\ref{xray_prop} is not correct, the true $N_{\rm H}$
of the source being $>10^{24}{\rm cm}^{-2}$. 
 There are two more Compton-thick sources (199 and 369) in our sample 
  which are transmission dominated. 
  In these the X-ray absorption turnover is redshifted at low energies
($5-6$\,keV) and an observer's frame $N_{\rm H}\sim 3\times 10^{23}{\rm cm}^{-2}$
is measured, which yields a rest-frame $N_{\rm H}>10^{24}{\rm cm}^{-2}$ when the
$(1+z)^{2.65}$ correction is applied.
  The most secure Compton-thick source is 369 for which 
    there is  spectroscopic redshift available and thus a more reliable
determination of the rest-frame column density.   
    Note that this source is not included in the CDFN Compton-thick sample
     of \citet{Georgantopoulos2009}, 
       because it has a flux lower than their adopted flux limit of
$10^{-15}$\,erg\,cm$^{-2}$\,s$^{-1}$ in the 2-10 keV band. 

\begin{table*}
\centering
\caption{X-ray properties}
\label{xray_prop}
\begin{tabular}{ccccccc}
\hline\hline
IAU & A03 & rest-frame $N_{\rm H}$     & $\Gamma$ & Counts & $\log(L_{\rm 2.0-10.0\,keV})$ \\
    &     & $\times 10^{22}{\rm cm}^{-2}$ &          &        & ${\rm erg\,s^{-1}}$           \\
\hline
CXO J123605.83+620838.0 &  73 & $<1.56$                    & $1.4^{+0.6}_{-0.4}$    & 175 & 44.00 \\
CXO J123608.61+621435.1 &  84 & $50$                       & 1.8                    & 20  & 43.61 \\
CXO J123613.02+621224.1 &  98 & $49.2_{-12.7}^{+17.1}$     & 1.8                    & 127 & 44.15 \\
CXO J123614.14+621017.7 & 100 & $121.3_{-24.3}^{+21.5}$    & 1.8                    & 56  & 44.07 \\
CXO J123615.83+621515.5 & 107 & $15.0_{-12.0}^{+20.2}$ *1  & $0.56^{+0.41}_{-0.31}$ & 211 & 44.08 \\
CXO J123621.07+621303.3 & 125 & $8.5_{-3.5}^{+4.3}$        & 1.8                    & 33  & 42.53 \\
CXO J123621.94+621603.8 & 129 & $115_{-37}^{+11}$          & 1.8                    & 64  & 44.03 \\
CXO J123623.66+621008.7 & 140 & $18.9_{-8.4}^{+7.2}$       & 1.8                    & 69  & 43.33 \\
CXO J123627.53+621218.0 & 151 & $26.8_{-19.9}^{+31.8}$     & 1.8                    & 52  & 43.92 \\
CXO J123628.78+621140.0 & 156 & $62.8_{-28.5}^{+7.9}$      & 1.8                    & 42  & 43.27 \\
CXO J123631.25+620957.3 & 167 & $<16.0$                    & 1.8                    & 35  & 43.11 \\
CXO J123634.48+620941.8 & 181 & $5.8_{-3.8}^{+5.1}$        & 1.8                    & 66  & 43.45 \\
CXO J123636.90+621320.0 & 196 & $<298$                     & 1.8                    & 31  & 43.86 \\
CXO J123637.26+620637.5 & 198 & $17.3_{-14.2}^{+6.1}$      & 1.8                    & 46  & 42.91 \\
CXO J123637.28+621014.2 & 199 & $480_{-460}^{+1070}$       & 1.8                    & 26  & 43.65 \\
CXO J123638.94+621041.5 & 204 & $13.2_{-6.8}^{+8.8}$       & 1.8                    & 78  & 43.46 \\
CXO J123639.65+620936.4 & 208 & $<25.8$                    & 1.8                    & 27  & 42.88 \\
CXO J123642.11+621331.6 & 220 & $<2.93$                    & 1.8                    & 44  & 43.01 \\
CXO J123647.94+621019.9 & 246 & $99.7_{-18.9}^{+12.4}$     & 1.8                    & 118 & 43.93 \\
CXO J123648.28+621456.2 & 250 & $1.60_{-0.60}^{+0.90}$ *2  & $2.0^{+0.7}_{-0.5}$    & 170 & -     \\
CXO J123656.56+621513.1 & 290 & $138_{-50}^{+114}$         & 1.8                    & 81  & 43.93 \\
CXO J123657.91+622128.6 & 299 & $7.05_{-2.94}^{+2.35}$     & $1.71^{+0.22}_{-0.23}$ & 787 & 44.72 \\
CXO J123658.74+621459.2 & 302 & $<159$                     & 1.8                    & 34  & 42.97 \\
CXO J123659.32+621833.0 & 307 & $103.8_{-38.4}^{+27.1}$    & 1.8                    & 67  & 44.21 \\
CXO J123701.62+621146.2 & 317 & $<10.2$                    & 1.8                    & 17  & 42.99 \\
CXO J123702.43+621926.1 & 321 & $16.37_{-2.58}^{+2.87}$    & $1.92^{+0.22}_{-0.28}$ & 433 & 44.10 \\
CXO J123703.85+621530.9 & 329 & $1.50_{-1.10}^{+1.50}$ *2  & 1.8                    & 28  & -     \\
CXO J123705.12+621634.8 & 335 & $67.3_{-21.1}^{+24.0}$     & 1.8                    & 125 & 44.35 \\
CXO J123712.09+621211.3 & 369 & $890_{-430}^{+540}$        & 1.8                    & 31  & 43.84 \\
CXO J123713.65+621545.2 & 372 & $1380_{-680}^{+1140}$      & 1.8                    & 40  & 44.38 \\
CXO J123713.84+621826.2 & 374 & $31.9_{-10.6}^{+13.1}$     & $2.16^{+0.48}_{-0.50}$ & 200 & 43.03 \\
CXO J123725.50+621707.3 & 417 & $<6.3$                     & 1.8                    & 155 & 43.58 \\
CXO J123733.98+621624.0 & 434 & $90_{-34}^{+43}$           & 1.8                    & 27  & 43.44 \\
CXO J123737.04+621834.4 & 445 & $17.13_{-2.81}^{+3.37}$    & $2.12^{+0.31}_{-0.40}$ & 492 & 44.06 \\
CXO J123750.22+621359.3 & 470 & $170_{-70}^{+100}$         & 1.8                    & 105 & 44.16 \\
\hline\hline
\end{tabular}
\begin{list}{}{}
\item *1: Column density is more likely $>10^{24}{\rm cm}^{-2}$ (see text).
\item *2: {\it Observer's frame} column density
\end{list}
\end{table*}

Next, we compare the mean X-ray spectral indices by coadding the individual
X-ray spectra in the total 0.3-8 keV band for four different sets of sources
(see Table \ref{coadd}). We fit a simple power-law model to the data. It is
evident that our sources present a very hard X-ray spectrum. The total spectrum
of all 35 sources has $\Gamma\approx 0.9$ which is harder than the coadded
spectrum of all the detected X-ray sources in this band
\citep[$\Gamma=1.4$,][]{Tozzi2001}. Note that the derived mean spectrum is
comparable to the spectrum derived by \citet{Fiore2008} and
\citet{Georgantopoulos2008} for Dust-Obscured-Galaxies i.e. sources defined as
having $f_{24\mu m}/f_R> 1000$; the coadded spectrum of DOGs in our sample has
$\Gamma=1.05$. The sources with IRAC-only detections are significantly harder
than those with optical {\it HST} counterparts. This difference cannot be attributed
to the fact that they have on average different redshifts and hence different
K-corrections, as it would work in opposite directions. The {\it HST} detected
sources are intrinsically softer than the ones completely lacking an optical
counterpart.
   
\begin{table}
\centering
\caption{Co-added X-ray spectral properties}
\label{coadd}
\begin{tabular}{ccc}
\hline\hline
Sample    & No & $\Gamma$       \\
\hline  
Total     & 35 & $0.87\pm 0.05$ \\
{\it HST} & 18 & $1.02\pm 0.07$ \\
IRAC-only & 17 & $0.56\pm 0.08$ \\
DOGs      & 17 & $1.05\pm 0.06$ \\
\hline\hline
\end{tabular}
\end{table} 
 
\section{Optical Properties}

\begin{figure}
  \resizebox{\hsize}{!}{\includegraphics{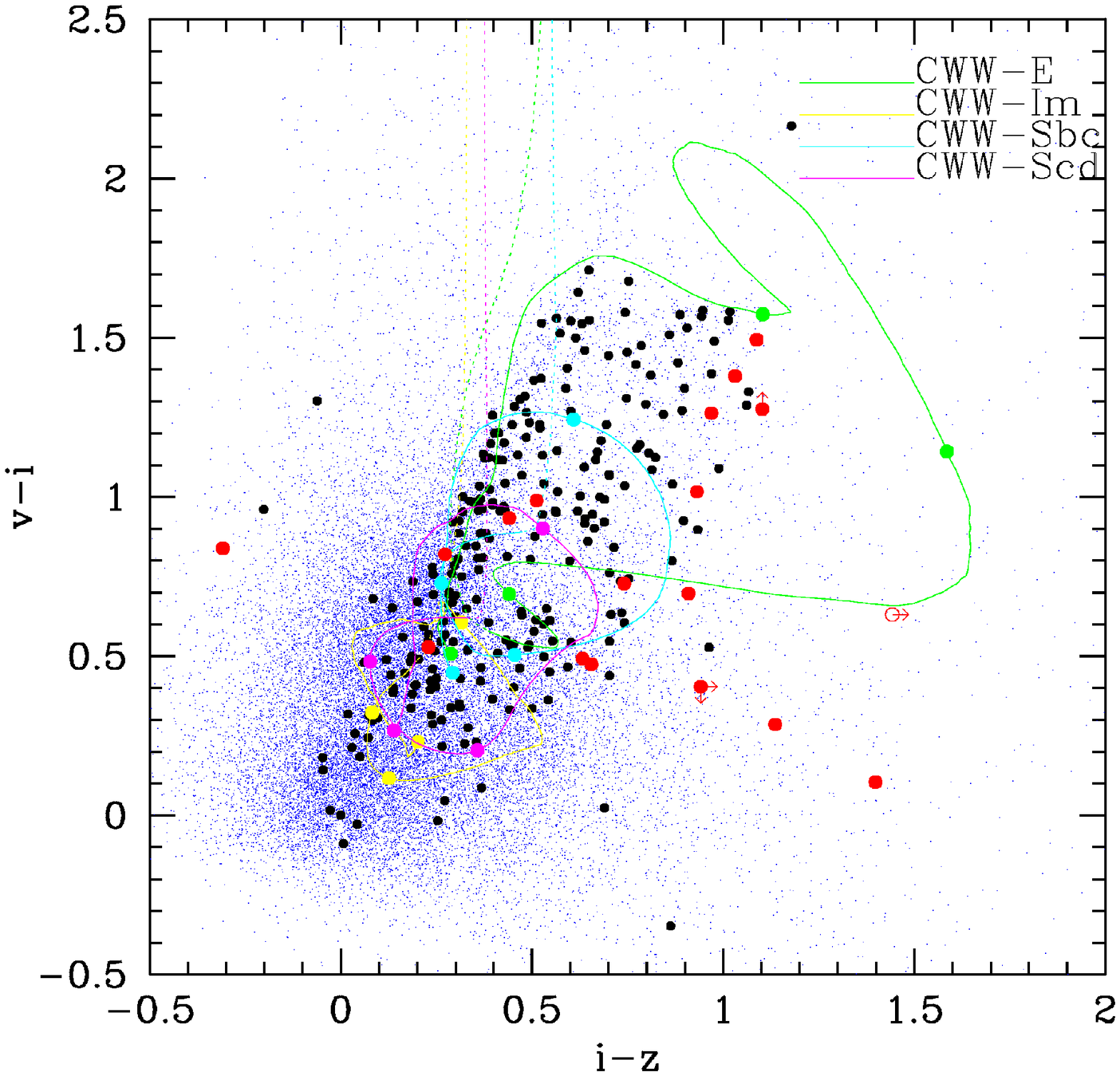}}
  \caption{Optical colours of {\it HST} detected sources. The optically faint
           sources of Tab.\,\ref{sample} appear in red and the X-ray sources
           appear in black. For comparison, all {\it HST} sources of
           \citet{Giavalisco2004} are plotted in blue. Colour lines track the
           colours of the \citet{Coleman1980} SED templates with redshift.}
  \label{HST_colours}
\end{figure}

In Fig\,\ref{HST_colours} we plot the $V_{606}-i_{775}$ vs $i_{775}-z_{850}$
colours of the 18 sources in our sample which have an ACS detection. Red points
mark the objects of our sample and black dots are {\it HST} sources associated with
an X-ray source. All the {\it HST} detections are shown in comparison in blue
points. The open circle with the right arrow represents source 317, which
has a low significance ($<2\sigma$) detection in the $V_{606}$ and $i_{775}$
bands and its $V_{606}-i_{775}$ colour cannot be accurately determined.
The coloured lines in Fig.\,\ref{HST_colours} trace the colours of various
templates with redshift with points marking $z=0,1,2,3$.
The templates are the \citet*{Coleman1980} galaxy templates for
elliptical, spiral, and irregular galaxies. Dotted lines correspond to
$z>4$, where the blue edge of the $V_{606}$ filter is redshifted to the
Lyman break (912\,$\AA$). The templates used are zeroed at shorter wavelengths
than the Lyman break, so the dotted lines are approximations of the colours
at high redshift. The sources of our sample appear significantly redder than
the overall X-ray population; 55.6\% (10/18) of them have $i_{775}-z_{850}>0.8$
compared to 14.4\% (42/291) of the overall X-ray population; the probability
that the $i_{775}-z_{850}$ colours of the red point being a random sample of the
black points in Fig\,\ref{HST_colours} is $<0.01\%$. The colours of the red
objects appear to be consistent with the elliptical template at $z=2\pm 1$,
which is backed up by the photometric redshifts of Tab.\,\ref{sample}. Red
optical colours are often associated with early-type morphologies
\citep[see][]{Bell2004}.

\begin{figure*}
  \resizebox{\hsize}{!}{\includegraphics{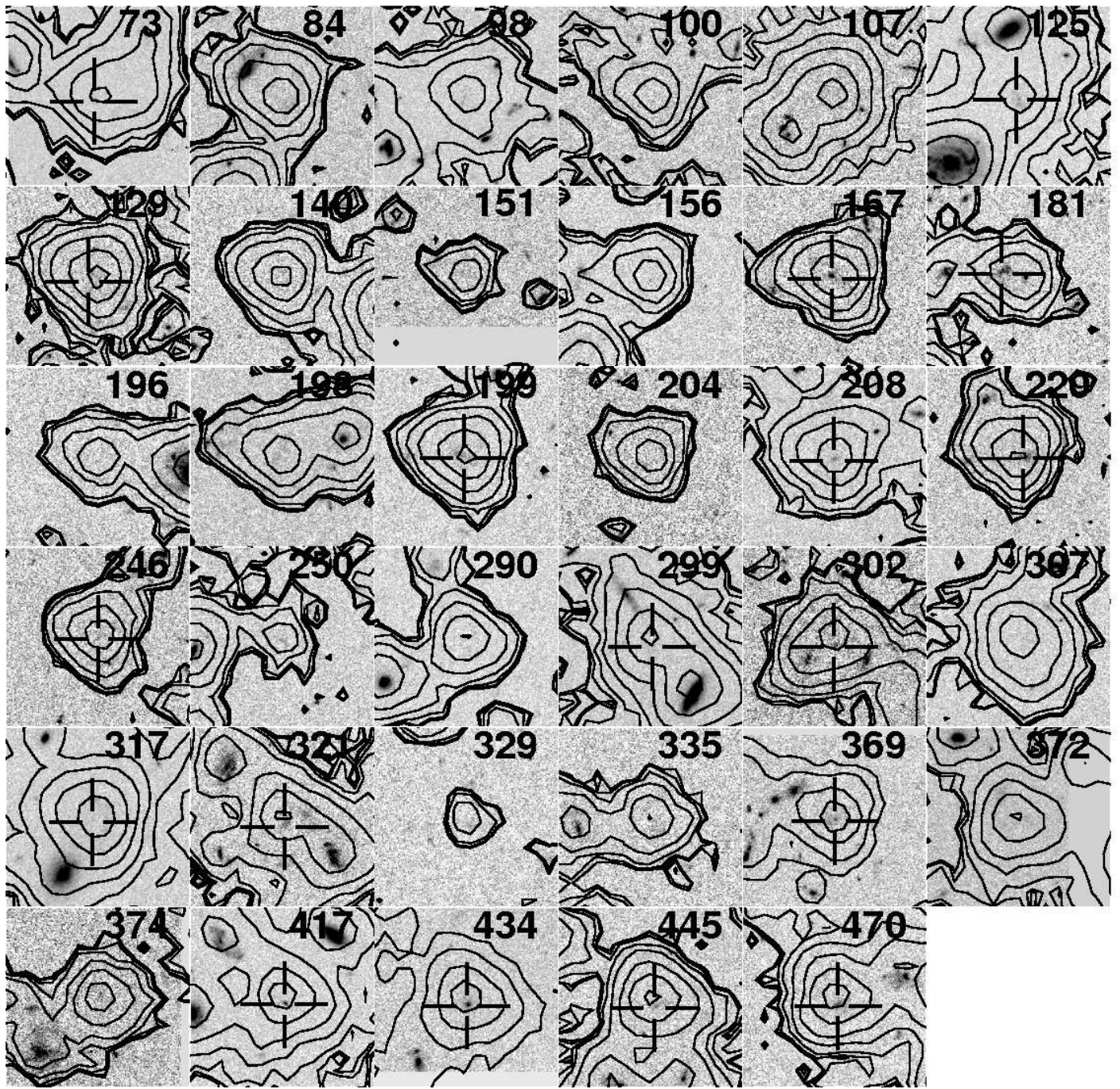}}
  \caption{{\it HST} cutouts of all sources listed in Tab.\,\ref{sample}. The
           images are combinations of the four ACS bands available to increase
           the signal-to-noise ratio, the contours represent the IRAC flux, and
           the crosses are centred on the GOODS source positions (if any) with
           an inner radius of 0.5\arcsec and an outer radius of 2\arcsec}
  \label{hst_cutouts}
\end{figure*}

In Fig.\,\ref{hst_cutouts} we plot
the {\it HST}-ACS cutouts of the galaxies of Tab.\,\ref{sample}. Each thumbnail
is a combination of the four ACS ($B-V-i-z$) images to increase the
signal-to-noise ratio and the contours represent the IRAC $3.6\,\mu{\rm m}$
flux, as in Fig.\,\ref{opt_positions}. The crosses mark the positions of the
{\it HST} sources with an inner radius of 0.5\arcsec and an outer radius of 2\arcsec.
We can see that for sources recovered with the {\it HST} the morphologies cannot be
determined.

Based on their near-infrared ($K_{\rm S}$) to optical colours, many of the
sources in our sample are associated with Extremely Red Objects
\citep[EROs;][]{Elston1988}. EROs are usually defined as galaxies having $R-K>5$
\citep[see also][for a selection based on the $I$-band with $I-K>4$]{Alexander2002}.
If we use the $R-$band limit ($R_{\rm AB}>26.5\Rightarrow R_{\rm Vega}>26.3$),
13/29 of the sources detected in $K_{\rm S}$ are EROs. However, the true optical
flux of the sources is fainter than this limit; if we use the $V_{606}$ ACS
magnitude for {\it HST} detected sources and
$V_{\rm AB}>27.8\Rightarrow V_{\rm Vega}>27.72$\footnote{The detection limit in
the $z_{850}$ ACS band is 27.8(AB) and according to the colours of {\it HST}
detections the $V_{606}$ limit is expected to be even higher.}, 28/29 of the
$K_{\rm S}$ detections have $V_{606}-K_{\rm S}>5$. Source 335 has
$V_{606}-K_{\rm S}>4.86$ and it is not detected with the {\it HST}, so it is highly
probable that it is too an ERO. Of the six sources not detected in
$K_{\rm S}$, 100, 151, 250, 329 and 374 are not detected with the {\it HST} either, so
they could be associated with EROs and 181 has $V_{606}-K_{\rm S}<5.03$.
Morphologically, EROs are mix of early and late-type systems
\citep{Cimatti2003,Gilbank2003} and the higher redshift ones tend to be more
late-type \citep{Moustakas2004}.

\section{Infrared properties}

\begin{figure}
  \resizebox{\hsize}{!}{\includegraphics{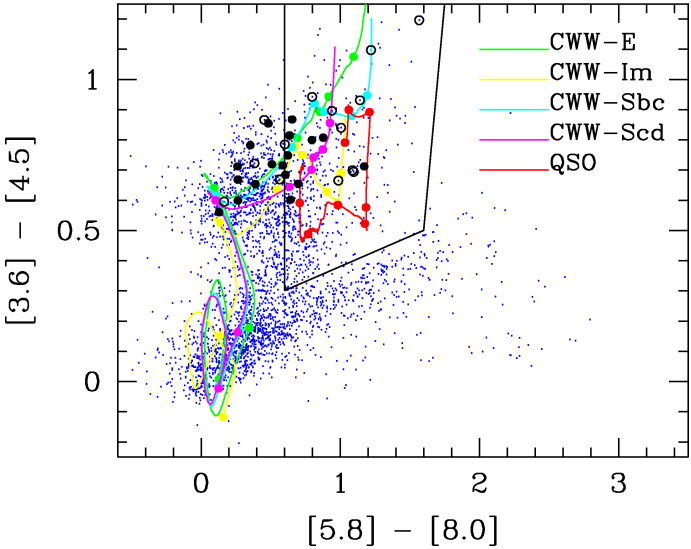}}
  \caption{Infrared colours of the sources listed in Tab.\,\ref{sample}. {\it HST}
           detected sources are plotted in filled circles and {\it HST}
           non-detections in open circles. All $3.6\,\mu{\rm m}$ detected
           sources are plotter in blue dots. The colour lines track the
           colours of the \citet{Coleman1980} SED templates with redshift
           and the black lines mark the region where infrared-selected AGN
           are located \citep{Stern2005}.}
  \label{IRAC_colours}
\end{figure}

We plot the mid-infrared colours of the sources of Tab.\,\ref{sample} in
Fig.\ref{IRAC_colours}. This diagram has been used by \citet{Stern2005} to
select AGN by their mid-infrared colours; a high fraction (up to 90\%) of
broad-line AGN is located inside the ``wedge'' marked with the black solid line
in Fig.\ref{IRAC_colours}. The {\it HST}-detected and non-detected sources of
Tab.\,\ref{sample} are plotted with filled and open circles respectively. The
coloured lines represent the colours of the \citet{Coleman1980} galaxy
templates as well as a QSO SED \citep{Elvis1994} for $0<z<8$, and points mark
$z=0,1,2,3,4,5,6,7$.

The sources of Tab.\,\ref{sample} are located both inside and outside the wedge
in Fig.\ref{IRAC_colours}, in the area where the red/optically-faint mid-IR AGN
of \citet{Georgantopoulos2008} lie. There is a separation in the colours of
{\it HST} detections and non-detections, the latter population having mid-infrared
colours mostly inside the wedge (10/14), and the rest being equally distributed
(9/18). This difference reflects on their [5.8]-[8.0] colours, a K-S test shows
that {\it HST} detections have bluer [5.8]-[8.0] colours than non-detections with
99.0\% ($2.6\,\sigma$) significance, and it is what causes the photometric
redshifts of {\it HST} non-detections to be higher. The galaxy templates of
\citet{Coleman1980} enter the wedge at $z=3$ so the photometric redshift of
those sources is $z>3$ if they are fitted with those templates. However, the
QSO template is in the wedge independent of the redshift because of its
power-law shape, in fact this is the reason why this diagram is a selection
method of mid-infrared AGN. So, a red [5.8]-[8.0] colour and the position
inside the wedge can be explained either with a QSO SED or a high redshift
($z>3$) galaxy SED.

\section{Discussion}

\subsection{High or intermediate-z sources ?}

In an earlier study of optically faint sources in the CDFN,
\citet{Alexander2001} concluded that they are moderately obscured AGN at
redshifts $z=1-3$, leaving a small margin for very high redshift QSOs in the
optically fainter subsample. However, their modest optical magnitude cutoff
($I>24$) includes sources that have ``normal'' X-ray to optical ratios
($f_{\rm x}/f_{\rm opt}\sim 1$). Studying more extreme cases based on
fainter optical fluxes or higher X-ray to optical ratios, \citet{Barger2003b}
and \citet{Koekemoer2004} suggest that the existence of very high redshift
objects cannot be definitely ruled out. In this study we compose a sample of
the optically faintest AGN ($R>26.5$) with robust infrared identifications
which is 83.3\% complete, including 35 of the 42 $R>26.5$ X-ray sources in the
common CDFN-GOODS-MOIRCS area used.

\subsubsection{Multi-$\lambda$ investigation}

For the 18 sources with an {\it HST} identification we have calculated their
photometric redshifts using up to 8 optical and infrared bands and the
results show that they are indeed at moderate redshifts, with a median
$z=2.61$, while all have $z<3.65$. The remaining 17 sources with no optical {\it HST}
detection present larger difficulty in constraining their redshift, and we have
attempted to derive photometric redshifts using the infrared bands. The
redshifts for sources with mid-infrared colours outside the ``wedge'' of
Fig.\,\ref{IRAC_colours}, or alternatively with
$f_{\nu\,5.8\,\mu{\rm m}}>f_{\nu\,8.0\,\mu{\rm m}}$ (6 sources) can be better
constrained and they have a median redshift of $z=2.39$. We assume that in these
cases the host galaxy dominates the mid-infrared colours which are well fitted
with non-AGN SEDs at $2<z<3$ (green, cyan, yellow, and magenta lines in
Fig.\,\ref{IRAC_colours}). The 9 sources with no optical {\it HST} detection and
$f_{\nu\,5.8\,\mu{\rm m}}<f_{\nu\,8.0\,\mu{\rm m}}$ have mid-infrared colours
compatible both with normal galaxy (or host galaxy) templates at high redshifts
or with QSO templates without a strong redshift constrain (see
\S\,\ref{photozs}); we try to constrain their redshifts using photometry from
lower energy infrared bands \citep[$24\,\mu{\rm m}$, see][]{Soifer2008}

\begin{figure}
  \resizebox{\hsize}{!}{\includegraphics{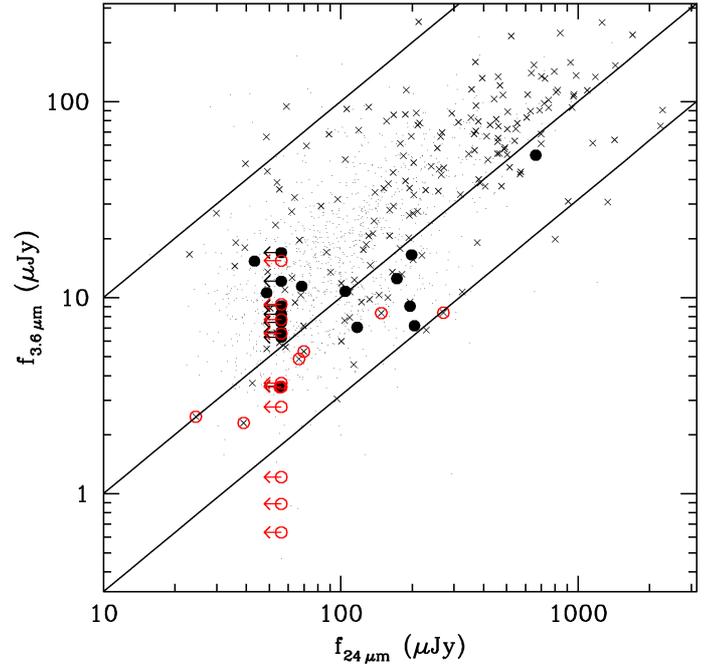}}
  \caption{$3.6\,\mu{\rm m}$ vs. $24\,\mu{\rm m}$ emission. Dots represent the
           overall $3.6\,\mu{\rm m}$ population with a $24\,\mu{\rm m}$ counterpart,
           crosses the counterparts of X-ray sources, black filled circles the
           sources of Tab.\,\ref{sample} detected with the {\it HST} and red open
           circles sources of Tab.\,\ref{sample} not detected with the {\it HST}. The
           lines mark $\log(f_{3.6\,\mu{\rm m}}/f_{24\,\mu{\rm m}})=0,-1,-1.5$.}
  \label{ir_plot}
\end{figure}

The position of the QSO templates inside the wedge is the result of the heating
of the circumnuclear dust with the radiation of the AGN, which results in a
red power-law SED ($f_{\nu}\propto\nu^{-\alpha}$ with $\alpha<-0.5$). The
circumnuclear dust can also extinguish the optical emission; \citet*{Dunlop2007}
claim that optically faint sources can be well fit with dusty and extremely
reddened ($A_V\simeq 4$) SEDs at moderate redshifts ($z\sim 2.5$).
We therefore check the mid-infrared $24\,\mu{\rm m}$ emission of the sources
of our sample to examine the dust properties. Of the 17 sources with no {\it HST}
detection, 7 are detected by MIPS, including 5/9 sources with
$f_{\nu\,5.8\,\mu{\rm m}}<f_{\nu\,8.0\,\mu{\rm m}}$. In Fig.\,\ref{ir_plot} we plot
the  $24\,\mu{\rm m}$ emission with respect to the $3.6\,\mu{\rm m}$ emission.
As dots we plot the overall population (non-AGN), as crosses we plot the
counterparts of the X-ray sources (AGN) and as black filled and red open
circles we plot the optically faint sources with and without an {\it HST} detection
respectively. We can see that the optically faint sources are in general fainter in
$3.6\,\mu{\rm m}$ emission with respect to their $24\,\mu{\rm m}$ emission,
which indicates dust extinction of optical and near-infrared wavelengths,
unless there is a strong $24\,\mu{\rm m}$ component from the AGN. The
$24\,\mu{\rm m}$ emission upper limits of the {\it HST} non detections can in most
cases be explained with dust obscuration, in cases where
$f_{3.6\,\mu{\rm m}}<5\,\mu{\rm Jy}$. The harder X-ray spectra of the {\it HST}
non-detections also suggest that (despite their somewhat higher redshifts)
these sources are more obscured, and this is what causes the faint optical
fluxes.

The four sources with no {\it HST} detection, $f_{5.8\,\mu{\rm m}}<f_{8.0\,\mu{\rm m}}$,
and no $24\,\mu{\rm m}$ detection (84, 151, 204, and 335) are possibly
associated with high redshift objects. However, the $24\,\mu{\rm m}$ upper
limits of three of them (151, 204, and 335) do not rule out dust obscuration
given their low $3.6\,\mu{\rm m}$ fluxes ($<3.6\,\mu{\rm Jy}$). Source 84 with
$f_{3.6\,\mu{\rm m}}=6.546\,\mu{\rm Jy}$ has an
$f_{3.6\,\mu{\rm m}}/f_{24\,\mu{\rm m}}$ ratio lower limit similar to the overall
AGN population with no clear signs of dust absorption and is therefore a high redshift
candidate. Its photometric redshift based on the 4 IRAC bands (4.50) is weakly
constrained and not reliable.

\subsubsection{Dropouts}

The Lyman-break technique \citep[e.g.][]{Steidel2003} is often used to select
high-redshift objects, and is based on the sudden drop in the flux of the
broad-band spectrum at wavelengths smaller than 912\,\AA.
\citet{Lehmer2005} used the {\it HST} observations to search for dropout sources
among the $z_{850}$-detected sources in the CDFN. They found two
$V_{606}$-dropouts at $z\gtrsim 5$ and one $i_{775}$-dropout at $z\gtrsim 6$.
One of the two $V_{606}$-dropouts \citep[source 247 in ][]{Alexander2003} is not
included in Tab.\,\ref{sample} because it is near a brighter optical source
(1.6\,arcsec separation) and the two are blended in the IRAC image. It has
a spectroscopically confirmed redshift of $z=5.186$ \citep{Barger2003}.
The other $V_{606}$-dropout (source 246) has a red $i_{775}-z_{850}$ colour
(1.1), like most of the sources in the {\it HST}-detected sample, and its 
$V_{606}-i_{775}$ colour is a lower limit ($V_{606}-i_{775}>1.275$ if we take the
$V_{606}$ $2\,\sigma$ limit), being compatible both with early-type templates
at $z\sim1.5$ and with very high redshift late-type templates, depending on the
true $V_{606}-i_{775}$ colour. This source is also observed spectroscopically,
but neither a redshift nor a spectral type could be derived
\citep{Barger2003,Trouille2008}. It has no MIPS $24\,\mu{\rm m}$ detection,
but its mid-infrared colours ($[5.8]-[8.0]=0.265$) argue against a high
redshift, they are explained by normal galaxy templates with $2<z<3$ agreeing
with the photometric redshift (2.80) derived using 7 optical and infrared bands.

The $i_{775}$-dropout of \citet{Lehmer2005} is
source 317. Its photometric redshift (3.25) is based in lower significance
$B_{435}$, $V_{606}$, and $i_{775}$ measurements and it is not reliable.
This the source with the highest $24\,\mu{\rm m}$ flux in our sample
($f_{24\,\mu{\rm m}}=664\,\mu{\rm Jy}$), which would yield a mid-infrared
luminosity in the order of $10^{26}\,{\rm W}\,{\rm Hz}^{-1}$ in the
$3.5\,\mu{\rm m}$ rest-frame band if the source was lying at $z\sim 6$.
Its low $f_{3.6\,\mu{\rm m}}/f_{24\,\mu{\rm m}}$ ratio however ($10^{-1.1}$)
implies dust obscuration according to Fig.\,\ref{ir_plot}, making it more likely
to be optically faint as a result of obscurarion rather than high redshift.

The dropout technique is proposed to select high redshift objects, it is
based however in 2-3 bands, usually including upper limits. All sources in
our sample are ``optical dropouts'' in the sense that they are not detected
in wavelengths shorter than a limit, in this case the $R$-band optical, and
taking the {\it HST} observations into account, almost half of them are $z$-dropouts.
When considering their multi-band properties however we see that a small
fraction of them (if any) can be high redshift ($z\gtrsim 6$) sources, as
there are other processes which can cause the faint optical fluxes. Especially
since we are examining AGN-hosting systems the SEDs are a combination of the
AGN and the host galaxy and a simple colour selection can be misleading.

\subsection{Obscured AGN ?}
 
The colour-colour diagram of Fig.\,\ref{HST_colours} shows that the sources
which are detected by the {\it HST} have on average redder optical colours than the
overall X-ray population and are compatible with elliptical galaxy SEDs.
\citet{Rovilos2007} and \citet{Georgakakis2008} have shown that red-cloud AGN
are obscured post-starburst systems. Instead, \citet{Brusa2009} attribute the
red colours to dust reddening rather than an evolved stellar population. The
fact however remains that X-ray AGN with red optical colours have a high
fraction of obscured sources.

A number of studies \citep[e.g.][]{Alexander2001,Mignoli2004,Koekemoer2004}
relate high X-ray-to-optical ratios to the Extremely Red Objects sample, with
$R-K>5$. The sources of Tab.\,\ref{sample} have by definition high
X-ray-to-optical ratios. Using the
{\it HST} magnitudes, we get $\log(f_{\rm x}/f_{\rm opt})>0.52$ between the ACS-$V$
band ($\lambda_{\rm eff}=6060\,\AA$) and the (0.5-10)\,keV X-ray band, assuming
an optical limit $V_{606}>27.8\,{\rm (AB)}$ for non detected sources (see
\S\ref{sample_selection} and Fig.\,\ref{fxfV}). They also have very red colours
with $V_{606}-K_{\rm S}>5$.  X-ray detected EROs are assumed to be
low-Eddington obscured AGN and they show on average high $f_{\rm x}/f_{\rm opt}$
values \citep{Brusa2005}, like our sources. The {\it HST}-detected subsample which
shows red optical colours have similar characteristics with the high X/O EROs
of \citet{Mignoli2004}, although this is a much fainter sample. Their sample
is detected in the $K$-band and has bulge-dominated morphologies, dominated by
their host galaxies \citep[see also][]{Maiolino2006}.

The sources in our sample are selected to have faint optical fluxes, and 
thus naturally many of them are associated with dust obscured galaxies 
(DOGs). DOGs \citep{Dey2008,Pope2008} are a class of objects, selected by their
high mid-infrared to optical fluxes ($f_{\nu}(24\,\mu{\rm m})/f_{\nu}(R)>1000$);
they are found at a redshift of $z\simeq 2$ \citep[e.g.][]{Pope2008} 
and are thought to be marking a phase of bulge and black-hole growth.
Hydrodynamic simulations \citep{Narayanan2009} suggest that DOGs are rapidly
evolving from starburst to AGN-dominated systems through mergers in cases where
$f_{\nu}(24\,\mu{\rm m})\gtrsim 300\,\mu{\rm Jy}$. Alternatively, in cases
where $f_{\nu}(24\,\mu{\rm m})\lesssim 300\,\mu{\rm Jy}$ the evolution is
secular, that is through smaller gravitational perturbations. Their
morphological characteristics \citep{Melbourne2009} reveal less concentrated
systems for lower luminosities.

Considering the $V_{606}$ magnitudes, and a $V_{606}>27.8$ in cases of
{\it HST} non-detections, all objects in Tab.\,\ref{sample} with a MIPS detection
(17/35 sources) have DOG characteristics\footnote{They all have
$f_{\nu}(24\,\mu{\rm m})/f_{\nu}(V_{606})>1000$, except source 196 with
$f_{\nu}(24\,\mu{\rm m})/f_{\nu}(V_{606})>890$}, and based on $24\,\mu{\rm m}$
upper limits the rest are not incompatible with DOGs, as the
$f_{\nu}(24\,\mu{\rm m})/f_{\nu}(V_{606})$ upper limit is always $>1250$.
Therefore half of our sample have characteristics consistent with DOGs and this
is only a lower limit. Only one source (317) has
$f_{\nu}(24\,\mu{\rm m})>300\,\mu{\rm Jy}$. Their red optical colours ({\it HST}
detections) are suggestive of early-type bulge-dominated morphologies
\citep{Bell2004,Mignoli2004}, although the morphologies cannot be determined
from direct observations, in line with their low $24\,\mu{\rm m}$ fluxes
\citep[see also][]{Bussmann2009}. In conjunction with the obscured nature of
the AGN, this means that the host galaxy is dominating the optical light.

On the basis of X-ray stacking analysis in deep X-ray fields,
\citet{Fiore2008,Fiore2009,Treister2009} propose that DOGs may be hosting a
large fraction of Compton-thick sources
\citep[see also][]{Georgantopoulos2008}. The sources of our sample appear to
be more obscured than the overall X-ray population, the average spectral index
being $\Gamma=0.87$ compared to $\Gamma=1.4$ \citep{Tozzi2001}. The stacked
spectrum of X-ray detected DOGs in the {\it Chandra} deep fields is
$\Gamma\simeq 0.7$ \citep{Georgantopoulos2008}, comparable to these of our
total and DOG samples. The average $\Gamma$ of our DOG sample is very close to
the value derived by Georgantopoulos et al. (2010, submitted) using all DOGs in
the CDFN regardless of their optical detection (or lack of it), a sample broadly
overlapping with ours.

According to Tab.\,\ref{xray_prop}, the average column density is a few times
$\rm 10^{23} cm^{-2}$. A few Compton-thick AGN may be also present in our
sample, this includes the reflection dominated source
\citep[107; see][]{Georgantopoulos2009}. Two more ``mildly'' Compton-thick sources 
i.e. with column densities $N_{\rm H} \sim 10^{24}$ $\rm cm^{-2}$ are directly 
identified on the basis of their absorption turnover entering the {\it Chandra}
passband. Finally, we note that a number of our sources, (10/35) of
Tab.\,\ref{xray_prop} have unobscured hard X-ray luminosities
$L_{\rm x}>10^{44}\,{\rm erg\,s}^{-1}$ and intrinsic
$N_{\rm H}>10^{23}\,{\rm cm}^{-2}$ making them members of the QSO2 class. It is
interesting that these sources do not reveal their QSO nature except in X-rays.

\subsection{X-ray Luminosity Function Incompleteness}

In the CDFs a number of X-ray sources lack an optical counterpart, even in the
deepest ground-based optical surveys. More specifically, within the common
CDFN - GOODS - MOIRCS area there are 35 sources which are not detected at a
magnitude limit of $R\sim 26.5$ and 17 sources not detected at
$z_{850}\sim 27.8$.
These populations represent 11.3\% and 5.5\% of the X-ray population detected
in this area. These numbers should be considered as lower limits, as there are
7 more X-ray sources with no optical counterpart not included in
Tab.\,\ref{sample} because either they are not detected with {\it Spitzer} or have
unreliable photometry. Moreover, there is a number of stars, normal galaxies,
or ultraluminous X-ray sources \citep{Hornschemeier2003,Bauer2004} among the
310 X-ray sources, so the fraction of non optically identified X-ray AGN is
even higher.

This sample of optically faint X-ray sources is not random in their
redshift distribution; the redshifts calculated in \S\,\ref{photozs} and the
K-S test performed show that the optically faint X-ray sources have
significantly higher redshifts than the overall population and the {\it HST}
non detections even higher. The redshifts listed in Tab.\,\ref{sample} are
in 34/35 cases higher than 1.5. If we count the sources within the common
GOODS - IRAC - MOIRCS area with an optical detection in \citet{Capak2004} and
$z>1.5$ their number is 41. So, the incompleteness of the X-ray $z>1.5$
population with an optical identification and a redshift estimation is
$\sim 50\%$ and it becomes even higher at higher redshifts; 29/35 of our
sources have $z>2$.

This has implications in the calculation of the X-ray luminosity function at
high redshift ($z>2-4$) and its evolution
\citep[e.g.][]{Silverman2008,Aird2008,Yencho2009,Aird2010}. More specifically,
this incompleteness affects the faint end of the XLF, as its ``knee'' at z=2 is
at $\log L_{\rm x}\sim 44.5\,{\rm erg}\,{\rm s}^{-1}$
\citep[$\log L_{*}=44.4-45.0\,{\rm erg\,s}^{-1}$ depending on the luminosity
evolution model, ][]{Silverman2008,Aird2010} and 34/35 of the sources in
Tab.\,\ref{xray_prop} have unobscured X-ray luminosities
$L_{\rm x}<10^{44.5}{\rm erg\,s}^{-1}$ in the $2-8$\,keV band.

\begin{figure}
  \resizebox{\hsize}{!}{\includegraphics{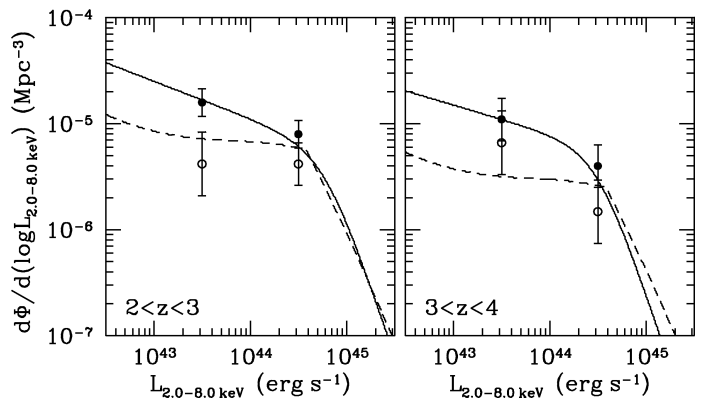}}
  \caption{The space density of AGN in the CDFN in two redshift bins
           ($2<z<3$, left, and $3<z<4$, right) and two luminosity bins
           ($10^{43}<L_{2.0-8.0\,{\rm keV}}<10^{44}$, and
           $10^{44}<L_{2.0-8.0\,{\rm keV}}<10^{45}$), with no correction applied
           for optical non-detections. Open symbols are based on the sources
           in the CDFN - GOODS - MOIRCS area with a redshift determination in
           \citet{Barger2003} and \citet{Trouille2008} and filled symbols are
           based on the combination of those sources with those of
           Tab.\,\ref{sample}. The solid and dashed lines represent the
           best-fit mod-PLE and LDDE models respectively
           \citep[see][]{Silverman2008}.}
  \label{lf_plot}
\end{figure}

In order to test how the inclusion of the sources in Tab.\,\ref{sample} can
affect the luminosity function at high redshift, we calculate the luminosity
function of AGN in two redshift bins, $2<z<3$, and $3<z<4$, with and without
their contribution, using the method described by \citet{Page2000}. We use the
combination of the CDFN catalogues of \citet{Barger2003} and
\citet{Trouille2008} and select sources in the combined CDFN - GOODS - MOIRCS
area. We do not make any distinction between spectroscopic and photometric
redshifts and we do not correct for optically unidentified sources. This has an
effect on the derived luminosity function only when we do not include the
sources of Tab.\,\ref{sample}, as the final sample is nearly complete.

The result is plotted in Fig.\ref{lf_plot}. We can see that the inclusion of the
new sources severely affects the calculated luminosity function. We note
that in most cases in the literature the incompleteness caused by optically non
detected sources is treated by, either taking an optically bright sample which is
spectroscopically complete, with the side effect of undersampling the high
redshift - low luminosity cases \citep[e.g.][]{Yencho2009}, or using
simplistic assumptions, like that their redshift distribution is directly
connected to their X-ray flux \citep[e.g.][]{Silverman2008}, which is generally
not true, or treating the incompleteness as a free parameter when fitting the
datapoints to determine the luminosity function evolution
\citep[e.g.][]{Aird2010}. In Fig.\ref{lf_plot} we also compare our datapoints
with models of
AGN evolution, which we adopt from \citet{Silverman2008}. These are a modified
Pure Luminosity Evolution (mod-PLE) model \citep*[see][]{Hopkins2007} (solid
line) and a Luminosity Dependent Density Evolution (LDDE) model (dashed line).
Our datapoints are remarkably close to the prediction of the mod-PLE model while
the non-corrected points are closer to those of the LDDE model. A thorough
investigation of the luminosity function evolution however would require taking
into account sources at all redshifts and luminosities which would vastly
increase the reliability of the statistics. This analysis is beyond the scope
of this work, but we caution the fact that the optically faint sources, despite
being a small minority of the X-ray sources can impact the LF in high redshifts
because of their redshift distribution.


\section{Conclusions}
We have examined the mid-IR and X-ray properties of 35 X-ray selected sources
in the GOODS-N area which are optically faint ($R_{AB} > 26.5$) and therefore
missed in ground-based optical observations. Instead of relying on previous
work on the matching between X-ray and optical counterparts, our sample has
been compiled anew. We find secure $3.6\mu{\rm m}$ counterparts for the X-ray
sources using a likelihood ratio technique and then in turn we search for 
their possible optical ground-based $R_{AB}$ counterparts. In case where there
are none down to $R_{AB}=26.5$, we search for {\it HST} ACS optical
counterparts. 18 sources have {\it HST} counterparts while the remaining have no
optical counterparts. Our findings can be summarized as follows:     
\begin{enumerate}
\item
Our sources populate moderate to high redshifts, being at markedly different
redshifts from the already spectroscopically identified population which peaks
at $z \sim 0.7$. In particular, the redshifts of the AGN with {\it HST} detections
have moderate values with a median redshift of 2.6. The redshifts of the
sources with IRAC detections only are definitely more uncertain; the objects
with blue [5.8]-[8.0] colours are probably located at redshifts comparable with
the {\it HST} population, $z\sim 2.5$, while the remaining sources could lie at
$z\gtrsim 3.2$.
A couple of $V$ and $i$ dropouts exist in our sample
\citep[previously reported by][]{Lehmer2005}, we however propose that they are
moderate redshift ($z=2-3$) dust-extinguished AGN, rather than lying at very
high redshift ($z>5$).
\item
The sources with no optical counterparts in deep ground-based optical surveys 
constitute a large fraction ($>50\%$) of the total source population at high
redshift $(z>2)$. This has important implications for the calculation and
modelling of the luminosity function at high redshift, which in the case of our
highly complete sample (97\% of the X-ray sources have spectroscopic or
photometric redshifts) is better represented by a modified PLE model.
\item
Our sources present very red colours. In particular, all 35 sources with
available $K_S$ magnitudes would be characterized as EROs on the basis of their 
$V_{606}-K_{\rm S}$ colour. 
\item
The mid-IR colours of the sources with {\it HST} counterparts lie outside the AGN
wedge, having blue [5.8]-[8.0] colours, in a region occupied by $z \sim 2 $
galaxies according to the galaxy templates. The majority {\it Spitzer} IRAC
mid-IR colours of the remaining sources with no optical counterparts lie within
the AGN 'wedge', suggesting either QSO templates or galaxy templates at high
redshift ($z>3$).
\item
We find 4 high redshift candidates based on their non detection with the {\it HST},
red $[5.8]-[8.0]$ colour, and non detection with MIPS at $24\,\mu{\rm m}$.
However, the low IRAC $3.6\,\mu{\rm m}$ of 3 of them do not definately rule out
dust obscuration in optical wavelengths.
\item 
17 out of 35 sources are detected in $24\,\mu{\rm m}$ and can be classified as
optically faint mid-IR bright galaxies. This class of objects is widely
believed to consist of reddened sources at moderate ($\sim 2$) redshifts.   
\item
The mean X-ray spectrum of our sources is very hard with $\Gamma \approx 0.9$,
much harder than the spectrum of all sources in the CDFs suggesting that we are
viewing heavily obscured sources. The X-ray spectroscopy on the individual
sources suggests that three sources are candidate Compton-thick AGN.          
    \end{enumerate}
    
Obviously the current deepest X-ray observations are not at par with the
present day optical spectroscopic capabilities. {\it Spitzer} has detected the
faintest X-ray sources and thus provided aid in the determination of their
properties and photometric redshifts.

\begin{acknowledgements}
IG acknowledges the receipt of a Marie Curie Fellowship Grant. The data used
here have been obtained from the {\it Chandra} X-ray archive, the NASA/IPAC Infrared
Science Archive and the MAST multimission archive at StScI.   
\end{acknowledgements}

\small{
\longtabL{1}{
\begin{landscape}
\begin{longtable}{ccccccccccccccccc}
\caption{\label{sample} List of CDFN X-ray sources which are correlated with a
         {\it Spitzer} $3.6\,\mu{\rm m}$ detection and lack an optical counterpart in
         \citet{Capak2004}.}\\
\hline\hline
A03 & $f_{(0.5-8.0)\,\rm{keV}}$ & RA (IRAC) & DEC (IRAC) & $B_{435}$ & $V_{606}$ & $i_{775}$ & $z_{850}$ & $HK'$ mag & $K_s$ mag & $f_{3.6\,\mu\rm{m}}$ & $f_{4.5\,\mu\rm{m}}$ & $f_{5.8\,\mu\rm{m}}$ & $f_{8.0\,\mu\rm{m}}$ & $f_{24\,\mu\rm{m}}$ & $\log\frac{f_{\rm x}}{f_{V_{606}}}$ & z-phot \\
    & erg\,cm$^{-2}$\,s$^{-1}$  & deg       & deg        & AB        & AB        & AB        & AB        & AB     & Vega   & $\mu$Jy              & $\mu$Jy              & $\mu$Jy              & $\mu$Jy              & $\mu$Jy             &                               &        \\
\hline
\endfirsthead
\caption{continued.}\\
\hline\hline
A03 & $f_{(0.5-8.0)\,\rm{keV}}$ & RA (IRAC) & DEC (IRAC) & $B_{435}$ & $V_{606}$ & $i_{775}$ & $z_{850}$ & $HK'$ mag & $K_s$ mag & $f_{3.6\,\mu\rm{m}}$ & $f_{4.5\,\mu\rm{m}}$ & $f_{5.8\,\mu\rm{m}}$ & $f_{8.0\,\mu\rm{m}}$ & $f_{24\,\mu\rm{m}}$ & $\log\frac{f_{\rm x}}{f_{V_{606}}}$ & z-phot \\
    & erg\,cm$^{-2}$\,s$^{-1}$  & deg       & deg        & AB        & AB        & AB        & AB        & AB     & Vega   & $\mu$Jy              & $\mu$Jy              & $\mu$Jy              & $\mu$Jy              & $\mu$Jy             &                               &        \\
\hline
\endhead
\hline
\endfoot
 73 & 3.65e-15 & 189.023735 & 62.144111 & $>$29.443 &    27.668 &    26.830 &    27.138 & 22.430 &    22.279 &  6.677 &  8.981 &   12.283 &   15.299 & $<$56.234 &    1.75 & 2.66 \\
 84 & 2.46e-16 & 189.035878 & 62.243257 &    -      &    -      &    -      & $>$27.800 & -      &    21.942 &  6.546 &  9.558 &   13.491 &   17.813 & $<$56.234 & $>$0.63 & 4.50 \\
 98 & 2.04e-15 & 189.054272 & 62.206709 &    -      &    -      &    -      & $>$27.800 & -      &    21.806 &  5.307 &  6.999 &    9.587 &    9.260 &    69.975 & $>$1.55 & 2.76 \\
100 & 9.39e-16 & 189.058928 & 62.171695 &    -      &    -      &    -      & $>$27.800 & -      & $>$21.693 &  2.293 &  3.458 &    4.470 &    7.094 &    39.019 & $>$1.21 & 3.95 \\
107 & 1.77e-15 & 189.065954 & 62.254337 &    -      &    -      &    -      & $>$27.800 & 21.842 &    20.805 & 15.406 & 19.171 &   25.913 & $<$3.548 & $<$56.234 & $>$1.49 & 3.13 \\
125 & 2.27e-16 & 189.088021 & 62.217674 &    27.885 &    27.687 &    26.960 &    26.218 & 21.569 &    21.479 &  8.205 & 9.576  &   13.275 &   10.552 & $<$56.234 &    0.55 & 1.58 \\
129 & 1.22e-15 & 189.091308 & 62.267668 & $>$30.195 &    30.080 & $>$29.676 &    28.734 & 21.938 &    20.930 & 10.754 & 13.262 &   14.610 &   10.325 &   104.577 &    2.24 & 2.78 \\
140 & 5.45e-16 & 189.098728 & 62.169285 &    -      &    -      &    -      & $>$27.800 & 22.355 &    21.321 &  8.326 & 10.362 &   12.848 &   10.158 &   148.205 & $>$0.98 & 2.28 \\
151 & 3.70e-16 & 189.114756 & 62.205093 &    -      &    -      &    -      & $>$27.800 & -      & $>$22.115 &  1.213 &  2.334 &    3.941 &    9.272 & $<$56.234 & $>$0.81 & 4.69 \\
156 & 3.27e-16 & 189.120125 & 62.194539 &    -      &    -      &    -      & $>$27.800 & -      &    21.622 &  3.661 &  4.338 &    5.364 &    5.004 & $<$56.234 & $>$0.75 & 2.39 \\
167 & 2.11e-16 & 189.130350 & 62.166161 &    28.476 &    27.716 &    27.188 &    26.958 & 22.388 &    20.763 &  9.035 & 10.847 &   14.562 &   14.136 &   196.049 &    0.53 & 3.02 \\
181 & 5.65e-16 & 189.144062 & 62.161763 &    27.080 &    27.299 &    26.825 &    26.169 & -      & $>$22.188 &  3.531 &  4.351 &    5.819 &    9.510 & $<$56.234 &    0.79 & 1.88 \\
196 & 4.07e-16 & 189.153696 & 62.222391 &    -      &    -      &    -      & $>$27.800 & -      &    22.433 &  2.468 &  3.420 &    5.149 &    7.215 &    24.525 & $>$0.85 & 3.48 \\
198 & 2.54e-16 & 189.155634 & 62.110729 &    -      &    -      &    -      & $>$27.800 & -      &    21.461 &  7.720 &  8.544 &   10.521 &    6.804 & $<$56.234 & $>$0.64 & 2.08 \\
199 & 2.39e-16 & 189.155645 & 62.170962 &    27.900 &    27.555 &    26.176 &    25.144 & 21.540 &    20.538 & 15.336 & 17.082 &   12.414 &   12.464 &    43.333 &    0.52 & 1.69 \\
204 & 5.42e-16 & 189.162390 & 62.178328 &    -      &    -      &    -      & $>$27.800 & -      &    21.979 &  3.503 &  4.134 &    4.721 &    6.503 & $<$56.234 & $>$0.97 & 2.78 \\
208 & 1.83e-16 & 189.165468 & 62.160321 & $>$29.074 &    28.308 &    27.292 &    26.360 & 21.554 &    20.820 &  9.069 & 10.075 &   10.004 &    7.076 & $<$56.234 &    0.71 & 2.37 \\
220 & 2.76e-16 & 189.175460 & 62.225511 &    28.457 &    27.620 &    26.800 &    26.527 & 22.693 &    21.213 &  7.178 &  8.872 &   12.096 &   11.517 &   204.792 &    0.61 & 2.72 \\
246 & 1.93e-15 & 189.199752 & 62.172370 & $>$29.451 & $>$29.629 &    28.354 &    27.251 & 21.841 &    21.499 &  6.267 &  7.419 &    6.570 &    4.656 & $<$56.234 & $>$2.26 & 2.80 \\
250 & 1.54e-15 & 189.201234 & 62.249071 &    -      &    -      &    -      & $>$27.800 & -      & $>$22.065 &  0.886 &  0.902 & $<$2.818 & $<$3.548 & $<$56.234 & $>$1.43 & -    \\
290 & 6.87e-16 & 189.235688 & 62.253708 &    -      &    -      &    -      & $>$27.800 & -      &    21.883 &  4.867 &  7.414 &   10.042 &   11.630 &    66.792 & $>$1.08 & 3.30 \\
299 & 5.44e-15 & 189.241374 & 62.358078 & $>$29.896 &    29.285 &    29.181 &    27.782 & 21.674 &    20.583 & 16.511 & 22.039 &   35.778 &   41.371 &   199.074 &    2.57 & 3.65 \\
302 & 1.20e-16 & 189.244707 & 62.249765 & $>$29.965 &    29.518 &    29.233 &    28.096 & -      &    21.955 &  7.050 &  9.907 &   10.557 &    9.140 &   117.399 &    1.01 & 2.40 \\
307 & 8.76e-16 & 189.247128 & 62.309081 &    -      &    -      &    -      & $>$27.800 & -      &    21.567 &  8.366 & 10.192 &   14.187 &   21.458 &   271.048 & $>$1.18 & 4.40 \\
317 & 9.01e-17 & 189.256662 & 62.196273 & $>$29.736 & $>$29.892 & $>$29.262 &    27.819 & 21.221 &    20.006 & 53.252 & 70.004 &   70.651 &   54.298 &   664.384 & $>$1.26 & 1.03 \\
321 & 3.97e-15 & 189.260124 & 62.323914 & $>$29.741 &    28.809 &    28.317 &    27.684 & -      &    21.656 &  7.486 & 10.640 &   12.880 &   13.057 & $<$56.234 &    2.24 & 2.55 \\
329 & 3.05e-16 & 189.266108 & 62.258594 &    -      &    -      &    -      & $>$27.800 & -      & $>$22.065 &  0.634 &  0.772 & $<$2.818 & $<$3.548 & $<$56.234 & $>$0.72 & -    \\
335 & 1.21e-15 & 189.271417 & 62.276409 &    -      &    -      &    -      & $>$27.800 & 22.249 &    22.861 &  2.770 &  3.357 &    4.201 &    6.402 & $<$56.234 & $>$1.32 & 4.60 \\
369 & 3.84e-16 & 189.300284 & 62.203440 &    29.292 &    28.546 &    27.613 &    27.172 & 21.945 &    20.728 & 10.584 & 14.332 &   21.049 &   20.979 &    48.774 &    1.12 & 2.80 \\
372 & 4.46e-16 & 189.306883 & 62.262653 &    -      &    -      &    -      & $>$27.800 & 22.156 &    22.282 &  9.228 & 13.102 &   19.188 &   16.181 & $<$56.234 & $>$0.89 & 4.90 \\
374 & 1.69e-15 & 189.307814 & 62.307393 &    -      &    -      &    -      & $>$27.800 & -      & $>$22.069 &  3.490 &  6.133 &    9.344 &   15.957 &    55.452 & $>$1.47 & 4.27 \\
417 & 7.35e-16 & 189.356457 & 62.285502 &    27.678 &    27.280 &    26.292 &    25.779 & 22.512 &    20.886 & 12.484 & 15.920 &   20.284 &   19.973 &   172.580 &    0.90 & 2.44 \\
434 & 8.21e-16 & 189.392075 & 62.273511 &    28.962 &    27.275 &    26.012 &    25.042 & 21.725 &    20.434 & 12.114 & 12.983 &    9.846 &    6.153 & $<$56.234 &    0.94 & 3.39 \\
445 & 4.43e-15 & 189.404382 & 62.309580 & $>$29.544 &    28.945 &    28.249 &    27.339 & 20.903 &    21.146 & 11.426 & 14.165 &   17.488 &   15.500 &    68.500 &    2.34 & 2.52 \\
470 & 1.95e-15 & 189.459284 & 62.233180 &    27.843 &    27.448 &    25.955 &    24.867 & 21.549 &    20.254 & 16.937 & 19.801 &   20.448 &   21.565 & $<$56.234 &    1.39 & 1.67 \\
\end{longtable}
\end{landscape}
}}

\end{document}